\documentclass[12pt,a4paper]{article}
\usepackage{a4wide}
\usepackage{latexsym}
\usepackage{epsf}
\usepackage{hyperref}
\usepackage{amssymb}
\begin{document}
\def\be{\begin{equation}}
\def\ee{\end{equation}}
\def\bea{\begin{eqnarray}}
\def\eea{\end{eqnarray}}
\def\pd{\partial}
\def\a{\alpha}
\def\b{\beta}
\def\g{\gamma}
\def\d{\delta}
\def\di{\mathrm{d}}
\def\k{{\bf k}}
\def\m{\mu}
\def\n{\nu}
\def \h{\mathcal{H}}
\def \hh{\mathcal{G}}
\def\t{\tau}
\def\th{\theta}
\def\p{\pi}
\def\th{\theta}
\def\l{\lambda}
\def\O{\Omega}
\def\r{\rho}
\def\s{\sigma}
\def\e{\epsilon}
  \def\scri{\mathcal{J}}
\def\cM{\mathcal{M}}
\def\tcM{\tilde{\mathcal{M}}}
\def\RR{\mathbb{R}}
\newcommand{\I}{\mathrm{i}}
\def\tr{\mathrm{tr}}
\def\D{\rm Diff}
\def\TD{\rm TDiff}
\def\WTD{\rm WTDiff}
\def\CD{\rm CDiff}
\def\A{A}


\hyphenation{re-pa-ra-me-tri-za-tion}
\hyphenation{trans-for-ma-tions}

\setlength\arraycolsep{1pt}


\begin{flushright}
IFT-UAM/CSIC-05-43\\
\end{flushright}

\vspace{1cm}

\begin{center}

{\bf\Large Transverse Fierz-Pauli symmetry.}
\\[1cm]
E.~\'Alvarez$^a$\, D.~Blas$^b$, J.~Garriga$^b$, E.~Verdaguer$^{b}$
\\
$^a${\it  Instituto de F\'{\i}sica Te\'orica UAM/CSIC, C-XVI,
and  \\
Departamento de F\'{\i}sica Te\'orica, C-XI,\\
  Universidad Aut\'onoma de Madrid
  E-28049-Madrid, Spain}\\
$^b${\it Departament de F\'isica Fonamental, Universitat de Barcelona,\\
Diagonal 647, 08028 Barcelona, Spain.} \vspace{.5cm} \vspace{.5cm}

\vskip .8cm
\end{center}
\abstract \setlength{\baselineskip}{.5cm} We consider some flat
space theories for spin 2 gravitons, with less invariance than
full diffeomorphisms. For the massless case, classical stability
and absence of ghosts require invariance under transverse
diffeomorphisms ($\TD$), $h_{\mu\nu} \mapsto h_{\mu\nu} + 2
\pd_{(\n}\xi_{\mu)}$, with $\pd_\m\xi^{\mu}=0$. Generic $\TD$
invariant theories contain a propagating scalar, which disappears
if the symmetry is enhanced in one of two ways. One possibility is
to consider full diffeomorphisms ($\D$). The other (which we
denote $\WTD$) adds a Weyl symmetry, by which the Lagrangian
becomes independent of the trace. The first possibility
corresponds to General Relativity, whereas the second corresponds
to ``unimodular" gravity (in a certain gauge). Phenomenologically,
both options are equally acceptable. For massive gravitons, the
situation is more restrictive. Up to field redefinitions,
classical stability and absence of ghosts lead directly to the
standard Fierz-Pauli Lagrangian. In this sense, the $\WTD$ theory
is more rigid against deformations than linearized GR, since a
mass term cannot be added without provoking the appearance of
ghosts.


%


\newpage

\setcounter{page}{1}
\setcounter{footnote}{1}
\setlength{\baselineskip}{.77cm}
\tableofcontents
\newpage
\section{Introduction}

It has long been known that, in theories containing a massless
spin 2 graviton propagating in flat space, unitarity requires
invariance under ``transverse" diffeomorphisms. The argument runs
as follows \cite{vanderBij}. Consider a graviton of momentum $k$
travelling in the $z$ direction. The ``little group" of Lorentz
transformations which leave $k$ invariant has three generators.
One of them, $I_z$, corresponds to rotations in the $x,y$ plane.
The other two, $I_{0x}$ and $I_{0y}$, correspond to transverse
boosts combined with rotations in the $x,z$ and $y,z$ plane\footnote{These rotations correct
for ``aberration" of $k$ under
the respective boosts.}. This little group is isomorphic to the
Euclidean group in 2 dimensions. The standard helicity
polarizations of the graviton\footnote{Up to overall
normalization, $ h^{+}\equiv dx\otimes dx-dy\otimes dy$ and
$h^\times\equiv dx\otimes dy+dy\otimes dx$.} $h^{+}\pm i
h^{\times}$, transform unitarily under $I_z$ (picking up phases
$\exp{\pm 2 i\theta}$ under rotations of angle $\theta$)
cfr.\cite{vanderBij,Alvarezz}.
 But unitary representations of
the non-compact ``translations" $I_{0i}$ would be infinite
dimensional, leading to an infinite number of polarizations for
given $k$. This catastrophic degeneracy is avoided by declaring
the equivalence of polarizations which are related to one another
by standard gauge transformation $h_{\mu\nu} \mapsto h_{\mu\nu} + 2
k_{(\mu} \xi_{\nu)}$. It can then be shown that the effect of
$I_{0i}$ on the standard helicity eigenstates is ``pure gauge", and
in this sense $I_{0i}$ act trivially, producing no new states of
momentum $k$. The interesting point, however, is that the trace
$h=\eta^{\mu\nu}h_{\mu\nu}$ is Lorentz invariant (and hence
invariant under $I_{0i}$), and therefore it is sufficient to
consider the equivalence under ``transverse" gauge transformations,
which don't affect the trace,
\begin{equation} k_\mu\xi^{\mu}=0. \end{equation}
These form a subgroup which we shall refer to as transverse
Fierz-Pauli symmetry, or transverse diffeomorphisms ($\TD$).

This paper is devoted to the study of some flat space theories
containing spin two, that have {\em less} invariance than the full
diffeomorphisms ($\D$) of the standard Fierz-Pauli Lagrangian. We
start our discussion, in Section II, with the most general Lorentz
invariant Lagrangian for a massless graviton. We show that a
simple requirement of classical stability and absence of ghosts leads directly to $\TD$
invariance (and no more invariance than that).

In the non-linear regime, $\TD$ symmetry is realized in the so called
{\em unimodular} theories. The best known example is Einstein' s
1919 theory (cf. \cite{Alvarezz} for a recent reference), which
corresponds to the traceless part of the usual Einstein's
equations. This theory can be obtained from an action which is
still generally covariant, but where the determinant of the metric
is not dynamical \cite{Unruh:1988in}.
 The equations of motion are obtained from a
restricted variational principle where \be \d \sqrt{|g|}=0.
\label{rest}\ee The trace-free equations enjoy the property that a
cosmological term in the matter Lagrangian is irrelevant.
Nevertheless, the trace of Einstein's equations can be recovered
with the help of the Bianchi identity, and then a cosmological
term reappears in the form of an integration constant.

Alternatively, we may start with a variational principle where $\d
\sqrt{|g|}$ is unrestricted, and with an action which is invariant
only under $\TD$ \cite{Buchmuller:1988wx}.
 Generically, this leads to scalar-tensor
theories, where the determinant of the metric plays the role of a
new scalar. As we shall see, this new degree of freedom can be
eliminated by imposing an additional Weyl symmetry (by which the
action becomes independent of the determinant of the metric).
Thus, Einstein's unimodular theory can be thought of as the theory
of massless spin two fields which is invariant under $\TD$ plus
certain Weyl transformations.\footnote{Also, in the absence of
this additional Weyl symmetry, the new scalar can be given a mass
(which may be expected from radiative corrections, since it is not
protected by any symmetry). If the mass is large enough, this
leads us back to the situation where the extra scalar does not
propagate at low energies, which is effectively equivalent to the
scenario described by Eq. (\ref{rest})}

Section III is devoted to the massive theory. Mass terms which
preserve $\TD$ would give mass to the new scalar, but not to the
spin 2 polarizations. Graviton mass terms necessarily break {\em
all} invariance under diffeomorphisms. For massive particles, the
little group is that of ordinary spatial rotations $O(3)$, which
leads to finite dimensional
 unitary representations without the need of invoking any gauge symmetry.
Naively, one might think that this would increase the
arbitrariness in the choice of the kinetic term. Nevertheless, as
we shall see, the only theory with massive gravitons which is free
from tachyons or ghosts is equivalent to the Fierz-Pauli theory,
where the kinetic term is invariant under $\D$ (not just $\TD$),
and the mass term is of the standard form
$m^2(h_{\mu\nu}h^{\mu\nu}-h^2)$ (see also \cite{van}). In Section
IV we consider the propagator for TDiff invariant theories, and
the coupling to conserved matter sources. We conclude in Section
V.

Throughout this paper we will follow the Landau-Lifshitz time-like
conventions; the $n$-dimensional flat metric in particular, reads
$\eta_{\mu\nu}=diag\,(1,-1,\ldots,-1)$. Lagrangians are written in
momentum space as well as in configuration space, depending on the
context. It is usually trivial to shift from one language to the
other.

\section{Massless theory}

Let us begin our discussion with the most general Lorentz
invariant local lagrangian for a free massless symmetric tensor field $h_{\m\n}$,
 \be \label{MA}
{\mathcal{L}}= {\mathcal{L}^{I}}+\beta\ {\mathcal{L}^{II}}+ a\
{\mathcal{L}^{III}}+b\ {\mathcal{L}^{IV}},\ee where we have
introduced \bea &&{\mathcal{L}}^I={1\over 4}\ \partial_\mu
h^{\nu\rho}\partial^\mu h_{\nu\rho}, \quad
{\mathcal{L}}^{II}=-{1\over 2}\
\partial_\mu h^{\mu\rho}\partial_\nu h^\nu_\rho, \nonumber\\
&&{\mathcal{L}}^{III}={1\over 2}\ \partial^\mu h\partial^\rho
h_{\mu\rho}, \quad {\mathcal{L}}^{IV}=-{1\over 4}\ \partial_\mu
h\partial^\mu h. \eea
The first term is strictly needed for the propagation of spin two
particles, and we give it the conventional normalization. Before
proceeding to the dynamical analysis, which will be done in
Subsection 2.4, it will be useful to consider the possible
symmetries of (\ref{MA}) according to the values of $\beta$, $a$
and $b$.

\subsection{$\TD$ and enhanced symmetries.}

 Under a general transformation of
the fields $h_{\mu\nu}\mapsto h_{\mu\nu}+\delta h_{\mu\nu}$, and
up to total derivatives, we have \bea
\delta {\mathcal{L}}^I&=&-{1\over 2} \delta h_{\mu\nu} \Box h^{\mu\nu}, \nonumber\\
\delta {\mathcal{L}}^{II}&=& \delta h_{\mu\nu}
\partial^\rho\partial^{(\mu} h_\rho^{\nu)},\nonumber\\
\delta {\mathcal{L}}^{III}&=&-{1\over 2}\Big(\delta h
\partial^\mu\partial^\nu h_{\mu\nu}+\delta h_{\mu\nu}
\partial^\mu\partial^\nu h\Big),\nonumber\\
\delta {\mathcal{L}}^{IV}&=& {1\over 2} \delta h \Box
h.\label{variIV} \eea
It follows that the combination \cite{Alvarezz} \be {\mathcal{L}}_{A}
\equiv {\mathcal{L}}^I + {\mathcal{L}}^{II} \label{simtfp}\ee is
invariant under restricted gauge transformations \be
 \delta h_{\mu\nu}
=2\partial_{(\mu} \xi_{\nu)}, \label{gauge} \ee with \be
\partial_\mu\xi^\mu=0.\label{ttra}
\ee  Since ${\mathcal{L}}^{III}$ and ${\mathcal{L}}^{IV}$ are
(separately) invariant under this symmetry, the most general $\TD$
invariant Lagrangian has $\beta=1$, and arbitrary coefficients $a$
and $b$: \be {\mathcal{L}}_{\TD} \equiv {\mathcal{L}}_A + a\
{\mathcal{L}}^{III} + b\ {\mathcal{L}}^{IV}. \label{tdl}\ee

An enhanced symmetry can be obtained by adjusting $a$ and $b$
appropriately. For instance, $a=b=1$ corresponds to the
Fierz-Pauli Lagrangian \cite{Fierz}, which is invariant under full
diffeomorphisms ($\D$), where the condition (\ref{ttra}) is
dropped. In fact, a one parameter family of Lagrangians can be
obtained from the Fierz-Pauli one through non-derivative field
redefinitions, \be \label{cov} h_{\m\n}\mapsto h_{\m\n}+\lambda h
\eta_{\m\n}, \quad\quad (\lambda\neq -1/n)\ee where $n$ is the
space-time dimension and the condition $\lambda\neq -1/n$ is
necessary for the transformation to be invertible. Under this
redefinition, the parameters in the Lagrangian (\ref{tdl}) change
as \be a\mapsto a+\lambda\left(an-2\right), \quad b\mapsto
b+2\lambda(nb-a-1)+\lambda^2(b n^2-n(2a+1)+2). \label{mas}\ee
Starting from $a=b=1$, the new parameters are related by \bea
\label{conftr} b&=&\frac{1-2 a+(n-1) a^2}{(n-2)}. \eea It follows
that Lagrangians where this relation is satisfied are equivalent
to Fierz-Pauli, with the exception of the case $a=2/n$, which
cannot be reached from $a=1$ with $\lambda\neq -1/n$.


A second possibility is to enhance $\TD$ with an additional Weyl
symmetry, \be\delta h_{\mu\nu} = {2\over n} \phi
\eta_{\mu\nu},\label{wesu}\ee by which the action becomes
independent of the trace. In the generic transverse Lagrangian
${\mathcal{L}}_{\TD}[h_ {\mu\nu}]$ of Eq. (\ref{tdl}), replace $h_
{\mu\nu}$ with the traceless part \be h_{\mu\nu}\mapsto\hat
h_{\mu\nu}\equiv h_{\mu\nu}-(h/n) \eta_{\mu\nu}\label{repl}.\ee
This is formally analogous to (\ref{cov}) with $\lambda=-1/n$, but
cannot be interpreted as a field redefinition. As such, it would
be singular, because the trace $h$ cannot be recovered from $\hat
h_{\mu\nu}$. The resulting Lagrangian \be
{\mathcal{L}}_{\WTD}[h_{\mu\nu}]\equiv{\mathcal{L}}_{\TD}[\hat
h_{\mu\nu}],\label{wtddef}\ee is still invariant under $\TD$ [the
replacement (\ref{repl}) does not change the coefficients in front
of the terms ${\mathcal{L}}^I$ and ${\mathcal{L}}^{II}$].
Moreover, it is invariant under (\ref{wesu}), since $\hat
h_{\mu\nu}$ is. Using (\ref{mas}) with $\lambda=-1/n$, we
immediately find that this ``$\WTD$" symmetry corresponds to
Lagrangian parameters \be a={2\over n},\quad \quad
 b = {n+2\over n^2}. \ee
This is the exceptional case mentioned at the end of the previous
paragraph. Note that the densitized metric $\tilde g_{\mu\nu} =
g^{-1/n} g_{\mu\nu} \approx \eta_{\mu\nu} + \hat h_{\mu\nu}$
enjoys the property that $\tilde g =1$. This is the starting point
for the non-linear generalization of the $\WTD$ invariant theory,
which is discussed in Subsection 2.5.

It is easy to show that $\D$ and $\WTD$ exhaust all possible
enhancements of $\TD$ for a Lagrangian of the form (\ref{MA}) (and
that, in fact, these are its largest possible gauge symmetry
groups). Note first, that the variation of ${\mathcal{L}}^I$
involves a term $\Box h^{\m\n}$.
 For arbitrary $h_{\mu\nu}$, this will only cancel against other
 terms in (\ref{variIV}) provided that the transformation is of the form
\be
\delta h_{\mu\nu} = 2 \partial_{(\mu} \xi_{\nu)}+{2\phi\over n}\eta_{\mu\nu},
\ee
for some $\xi^{\mu}$ and $\phi$. The vector can be decomposed as
\be
\xi_{\mu}=\eta_{\mu} +\pd_{\mu} \psi
\ee
where
$\partial_{\mu}\eta^{\mu}=0$. Using (\ref{variIV}) we readily find
\bea
\delta{\mathcal L}&=&
\eta_\nu (\beta-1)\Box(\partial_\mu h^{\mu\nu})\nonumber\\
&+&{\psi\over 2}\left[(b-a)\Box^2 h+( 2\beta -a-1)\Box(\partial_\mu\partial_\nu h^{\mu\nu})\right]\nonumber\\
&+&{\phi\over n}\left[(bn-a-1)\Box h + (2\beta-na)
\partial_\mu\partial_\nu h^{\mu\nu}\right]. \eea $\TD$ corresponds
to taking $\beta=1$, with arbitrary transverse $\eta^\mu$ and with
$\phi=\psi=0$. This symmetry can be enhanced with nonvanishing
$\phi$ and $\psi$ satisfying the relation \be n (a-1) \Box \psi =
2 (2-an) \phi,\label{relat1} \ee provided that \be
 b=\frac{1-2 a+(n-1) a^2}{(n-2)}.\label{relat2}
\ee Eq. (\ref{relat1}) ensures the cancellation of the terms with
$\partial_\mu\partial_\nu h^{\mu\nu}$, and Eq. (\ref{relat2})
eliminates  terms containing the trace $h$. Eq. (\ref{relat2})
agrees with (\ref{conftr}), and therefore the Lagrangian with the
enhanced symmetry is equivalent to Fierz-Pauli, unless $a=2/n$,
which corresponds to $\WTD$\footnote{Incidentally, it may be noted
that for $n=2$ both possibilities coincide, since in this case the
symmetry of the Fierz-Pauli Lagrangian is full diffeomorphisms
plus Weyl transformations.}.

\subsection{Comparing $\D$ and $\WTD$}

Let us briefly consider the  differences between the two enhanced
symmetry groups. A first question is whether the Fierz-Pauli
theory ${\cal L}_{\D}$ is classically equivalent to ${\cal
L}_{\WTD}$. Since $\D$ includes $\TD$, we can use (\ref{wtddef})
to obtain \be {\delta {\cal S}_{\WTD}[h]\over \delta h_{\mu\nu}} =
{\delta {\cal S}_{\D}[\hat h]\over \delta \hat h_{\rho\sigma}}\
\left(\delta^\mu_{(\rho}\delta^\nu_{\sigma)} -{1\over
n}\eta_{\rho\sigma}\eta^{\mu\nu}\right).\label{eomrel}\ee Hence,
the $\WTD$ equations of motion are traceless $$ {\delta {\cal
S}_{\WTD}[h]\over \delta h_{\mu\nu}}\eta_{\mu\nu}\equiv 0.$$ In
the $\WTD$ theory, the trace of $h$ can be changed arbitrarily by
a Weyl transformation, and we can always go to the gauge where
$h=0$. Likewise, in the familiar $\D$ theory we can choose a gauge
where $h=0$. Then, $h_{\mu\nu}=\hat h_{\mu\nu}$, and the $\WTD$
equations of motion (e.o.m.) are just the traceless part of the
Fierz-Pauli e.o.m. Differentiating Eq. (\ref{eomrel}) with respect
to $x^\m$ and using the Bianchi identity
$$\partial_\rho\left({\delta {\cal S}_{\D}[h]\over \delta
h_{\rho\sigma}}\right)=0,$$ one easily
finds that ${\delta {\cal S}_{\WTD}[h]/ \delta h_{\mu\nu}}=0$
implies
$${\delta {\cal S}_{\D}[h]\over \delta
h_{\rho\sigma}}\ \eta_{\rho\sigma}= \Lambda.
$$
Hence, the trace of the Fierz-Pauli e.o.m. is also recovered from
the $\WTD$ e.o.m. (in the gauge $h=0$), up to an arbitrary
integration constant $\Lambda$ which plays the role of a cosmological
constant\footnote{Here we assume $\Lambda = O(h)$. }
. Thus, the two theories are closely related, but they are not quite the same.

Let us now consider the relation between the corresponding
symmetry groups. Acting infinitesimally on $h_{\m\n}$ they give
\bea
\delta^{D} h_{\m\n}&=&2 \partial_{(\m}\xi_{\n)}=2\partial_{(\m}\eta_{\n)}+\pd_{\m}
\pd_{\n}\psi\label{feq}\\
\delta^{WTD} h_{\m\n}&=&2
\partial_{(\m}\bar\eta_{\n)}+\frac{2}{n}\phi\eta_{\m\n}\label{seq} \eea where
$\partial_\m \eta^\m=\partial_\m \bar\eta^\m=0$. In (\ref{feq}) we have
 decomposed $\xi_{\n}=\eta_\nu + \pd_\nu\psi$ into transverse and longitudinal part.
 The intersection of $\D$ and $\WTD$ can be found by equating (\ref{feq}) and (\ref{seq})
\be 2\partial_{(\m}\eta_{\n)}+\pd_{\m} \pd_{\n}\psi=2
\partial_{(\m}\bar\eta_{\n)}+\frac{2}{n}\phi\eta_{\m\n}.\label{mac}
\ee
Taking the trace, we have
\be
\Box \psi=2\phi.\label{mac2}
\ee
The divergence of (\ref{mac}) now yields
\be
\Box (\bar\eta_\mu-\eta_\mu) = {n-1\over n} \Box \pd_\mu\psi. \label{wac}
\ee
Taking the divergence once more, we have
\be
\Box \phi=0.
\ee
Taking the derivative of (\ref{wac}) with respect to $\nu$, symmetrizing with respect to $\mu$ and $\nu$,
and using (\ref{mac}) and (\ref{mac2}),
we have $(n-2)\pd_{\m}
\pd_{\n}\Box \psi=0$. For $n\neq 2$ this implies $\pd_{\m}
\pd_{\n}\phi=0$, {\em i.e.}
$$
\phi=b_\mu x^{\mu} + c,
$$
where $b_\mu$ and $c$ are constants. Hence, not every Weyl transformation belongs to $\D$, since only the $\phi$'s which are linear in $x^{\mu}$ qualify as such. Conversely, the subset of $\D$ which can be expressed as Weyl transformations are the solutions of the conformal Killing equation for the Minkowski metric \cite{waldbook},
\be
\pd_{(\m}\xi^{CD}_{\n)}=\frac{1}{n}\phi  \eta_{\m\n},\label{confd}
\ee
where $\phi= \pd^\r \xi^{CD}_\r$ (and, as shown above, $\phi$ has to be a linear function of $x^{\mu}$). These solutions generate the so called conformal group, which we may denote by $\CD$. In conclusion,
 the enhanced symmetry groups $\D$ and $\WTD$ are not subsets of each other. Rather, their intersection is the set of $\TD$ plus $\CD$.

\subsection{Traceless Fierz-Pauli and $\WTD$}

An alternative route to the $\WTD$ invariant theory is to try and construct a
Lagrangian which will yield the traceless part of Einstein's equations.

It is clear, however, that we can only obtain traceless equations of motion from
 a Lagrangian which is invariant under Weyl transformations. If the e.o.m. are traceless,
 then $\delta S=0$ for variations of the form for $\delta h_{\mu\nu} \propto \eta_{\mu\nu}$.
 This symmetry is not included in $\D$, and therefore the traceless part of Einstein's equations
 cannot be recovered from this Lagrangian in every gauge. Rather, we should look for
 a Lagrangian which will yield the traceless part of Einstein's equations in {\em some} gauge.

Let us consider the $\D$ e.o.m. in momentum space
\be
 {\delta {\cal S}_{\D}[h]\over \delta  h_{\rho\sigma}}=K^{\r\s\m\n}_{\D} h_{\m\n},
\ee
where
\bea
&&8 K_{\D}^{\m\n\r\s}=k^2\left(\eta^{\m\rho}\eta^{\n\sigma}+\eta^{\m\sigma}\eta^{\n\rho}-
2\eta^{\m\n}\eta^{\rho\sigma}\right)-\nonumber\\
&&\left(k^{\m}k^{\rho}\eta^{\n\sigma}+ k^{\n}k^{\sigma}\eta^{\m\rho}+
k^{\m}k^{\sigma}\eta^{\n\rho}+ k^{\n}k^{\rho}\eta^{\m\sigma}-2 k^{\m}k^{\n}\eta^{\rho\sigma}
-2 k^{\rho}k^{\sigma}\eta^{\m\n}\right).
\eea
We can also define the traces
\bea
\tr K^{\m\n}_{\D}&=&\eta_{\r\s} K^{\r\s\m\n}_{\D}=
\frac{n-2}{4}\left(
k_{\rho}k_{\sigma}-k^2 \eta_{\rho\sigma}\right), \nonumber\\
 \tr \
\tr K_{\D}&=&\eta_{\m\n}\eta_{\r\s} K^{\r\s\m\n}_{\D}=-\frac{(n-1)(n-2)}{4}k^2.
\eea
The traceless part of the $K^{\r\s\m\n}_{\D}$,
\bea
&&8K_{\D}^t=8\left(K_{\D}-\frac{1}{n}\eta^{\m\n}\tr\,K_{\D}^{\rho\sigma}\right),
\eea
cannot be derived from a Lagrangian as it is not symmetric in the indices $(\r\s)$ vs. $(\m\n)$. Nevertheless,
we can still define traceless symmetric Lagrangians. One might think of substituting $\eta^{\m\n}$
in the previous expression by $\tr\,K_{\D}^{\m\n}$, and dividing
by its trace. However, this would be nonlocal.

For a local Lagrangian which is still invariant under $\TD$, we must restrict to
deformations which correspond to changes in the parameters $a$ and $b$ in (\ref{MA}).
The most general symmetric Lagrangian with these properties is of the form
\be
K_{t\D}^{\m\n\rho\sigma}\equiv K_{\D}^{\m\n\rho\sigma}-\eta^{\m\n}M^{\rho\sigma}-M^{\m\n}
\eta^{\rho\sigma},
\ee
with $M_{\r\s}$ a symmetric operator at most quadratic in the momentum.
Asking that the result be traceless leads to:
\be
M^{\m\n}=\frac{1}{n}\left(\tr\, K_{\D}^{\m\n}-(\tr\, M)\eta^{\m\n}\right),
\ee
which implies
\be
\tr\,M =\frac{1}{2n} \tr\,\tr\,K_{\D}.
\ee
Therefore
\be
M^{\m\n}=\frac{1}{n}\left(\tr\,K_{\D}^{\m\n}-\frac{1}{2n}(\tr\, \tr\,K_{\D})\eta^{\m\n}\right),
\ee
and we can write
\bea
8 K_{t\D}^{\m\n\rho\sigma}=
&&k^2\left(\eta_{\m\rho}\eta_{\n\sigma}+\eta_{\m\sigma}\eta_{\n\rho}\right)-
\left(k_{\m}k_{\rho}\eta_{\n\sigma}+ k_{\n}k_{\sigma}\eta_{\m\rho}+
k_{\m}k_{\sigma}\eta_{\n\rho}+ k_{\n}k_{\rho}\eta_{\m\sigma}\right)\nonumber\\
&&\hspace{1cm}-\frac{2(n+2)}{n^2}k^2\eta_{\m\n}\eta_{\rho\sigma}+\frac{4}{n}(
k_{\m}k_{\n}\eta_{\rho\sigma}+
k_{\rho}k_{\sigma}\eta_{\m\n}).
\eea
Moving back to the position space, this corresponds to the $\WTD$ Lagrangian, i.e.
the case $a=\frac{2}{n}$ and $b=\frac{n+2}{n^2}$ in (\ref{tdl}).
As shown before, this yields the traceless part of the Fierz-Pauli e.o.m.
in the gauge $h=0$.

A similar
analysis could be done for the massive case. However, as we shall see in the next section,
the corresponding Lagrangian has a ghost.

\subsection{Dynamical analysis of the general massless Lagrangian.}

The little group argument mentioned in the introduction indicates
that the quantum theory is not unitary unless the Lagrangian is
invariant under $\TD$. In fact, in the absence of $\TD$ symmetry
 the Hamiltonian is unbounded below. This leads to pathologies such as classical
 instabilities or the existence of ghosts.

To show this, as well as to analyze the physical degrees of freedom of the general massless theory (\ref{MA}),
 it is very convenient to use the
``cosmological" decomposition in terms of scalars, vectors, and
tensors under spatial rotations $SO(3)$ (see e.g. \cite{Mukhanov:1990me}),
\bea
h_{00}&=&\A \nonumber\\
h_{0i}&=& \pd_i B+V_i\nonumber\\
h_{ij}&=&\psi\delta_{ij}+\pd_i \pd_j E+2\pd_{(i}F_{j)}+t_{ij}
\label{cosdec}\eea where $\pd^iF_i= \pd^iV_i=\pd^it_{ij}=t^i_i=0$.
The point of this decomposition is that in the linearized theory
the scalars ($\A,B,\psi,E$), vectors ($V_i,F_i$) and tensors
($t_{ij}$) decouple from each other. Also, we can easily identify
the physical degrees of freedom without having to fix a gauge (see
Appendix A).

The tensors $t_{ij}$ only contribute to ${\mathcal L}^I$, and one
readily finds \be {}^{(t)}{\mathcal L}=-\frac{1}{4}t^{ij} \Box
t_{ij}\label{tensors}\ee The vectors contribute both to ${\mathcal
L}^I$ and ${\mathcal L}^{II}$. Working in Fourier space for the spatial coordinates and
after some straightforward algebra, we have
\be \label{vectors} {}^{(v)}{\mathcal{L}}=\frac{1}{2}
\k^2\left(V^i-\dot F^i\right)^2 + {1\over 2}(\beta-1)\left(\k^2F^i+\dot
V^i\right)^2.\label{vecs} \ee For $\beta=1$, corresponding to $\TD$
symmetry, there are no derivatives of $V^i$ in the Lagrangian.
Variation with respect to $V^i$ leads to the constraint $V^i-\dot
F^i=0$, which upon substitution in (\ref{vecs}) shows that there
is no vector dynamics.

Other values of $\beta$ lead to pathologies.  The Hamiltonian is
given by \be {}^{(v)}{\mathcal H}={(\Pi_F+\k^2 V)^2 \over 2\k^2}-{[\Pi_V
+(1-\beta)\k^2 F]^2\over 2(1-\beta)}  +{(1-\beta)\k^4 F^2\over 2}-
{\k^2 V^2\over 2},\label{ham} \ee where the momenta are given by
$\Pi_F=\k^2\left(\dot F-V\right)$ and $\Pi_V=(\beta-1)\left(\k^2 F+\dot V\right)$,
 and
we have suppressed the index $i$ in the vectors $F$ and $V$.
Because of the alternating signs in Eq. (\ref{ham}), the
Hamiltonian is not bounded below. Generically this leads to a
classical instability. The momenta satisfy the equations
$\dot\Pi_F = \k^2 \Pi_V$ and $\dot\Pi_V = -\Pi_F$. These have the
general oscillatory solution
$$
|\k| \Pi_V+\I\ \Pi_F= C \exp{\I(|\k| t +\phi_0)},
$$
where $C$ and $\phi_0$ are real integration constants. On
the other hand, $V$ and $F$ satisfy \bea
\ddot V+\k^2 V= {-\beta\over (\beta-1)} \Pi_F,\\
\ddot F+\k^2 F= {\beta \over (\beta-1)} \Pi_V.
\eea
For $\beta\neq 0$ these are equations for forced oscillators. For large times,
the homogeneous solution becomes irrelevant and we have
$$
V + \I|\k| F \sim \left({\beta C t\over (\beta-1) |\k|}\right)
\exp{\I(|\k| t + \phi_0)},
$$
whose amplitude grows without bound, linearly with time.
This classical instability is not present for $\beta=0$. However, in this case $F$ and $V$ decouple and we have
$$
 {}^{(v)}{\mathcal{L}}_{\beta=0}=\frac{1}{2} \k^2(\partial_\mu
F^i)^2 - {1\over 2}(\partial_\mu V^i)^2,
$$
so $V_i$ are ghosts.

Hence, the only case where the vector Lagrangian is not problematic is $\beta=1$,
corresponding to invariance under $\TD$. The scalar Lagrangian is then given by\footnote{
The equivalent expression in terms of gauge invariant combinations is given in Appendix A.}
\bea
{}^{(s)}{\mathcal{L}}_{\TD}&=&{1\over 4}\left[ (\partial_\mu \A)^2-2\k^2(\partial_\mu B)^2+N(\partial_\mu\psi)^2-2\k^2\partial_\mu\psi\partial^\mu E+\k^4(\partial_\mu E)^2 \right]\nonumber\\
&-&{1\over 2}\left[(\dot \A + \k^2 B)^2-\k^2 \dot B^2-\k^2\psi^2+2\k^4 E\psi-\k^6E^2+2\k^2\dot B(\psi-\k^2 E)\right]\nonumber\\
&+&{a\over 2}\left[(\dot \A-N\dot\psi+\k^2 \dot E)(\dot \A+\k^2 B)-\k^2(\A-N\psi+\k^2 E)(\dot B-\psi+\k^2 E)\right]\nonumber\\
&-&{b\over 4}\left[\partial_\mu(\A-N \psi +\k^2
E)\right]^2,\label{scalax} \eea where $N=n-1$ is the dimension of
space. It is easy to check that $B$ is a Lagrange multiplier,
leading to the constraint \be (N-1)\psi=(a-1) h, \ee where
$h=\A-N\psi+\k^2E$ is the trace of the metric perturbation.
Substituting this back into the scalar action (\ref{scalax}) we
readily find \be {}^{(s)}{\mathcal{L}}_{\TD}= -{\Delta b\over
4}(\partial_\mu h)^2, \ee where \be
 \Delta b\equiv b-\frac{1-2a+(n-1)a^2}{n-2}. \label{deltab0}
\ee Hence, the scalar sector contains a single physical degree of
freedom, proportional to the trace. Whether this scalar is a ghost
or not is determined by the parameters $a$ and $b$. For
$b=(1-2a+(n-1)a^2)/(n-2)$, corresponding to the enhanced
symmetries which we studied in the previous subsection, the scalar
sector disappears completely, and we are just left with the tensor
modes.
\subsection{Nonlinear theory}

Non-linear generalizations of $\TD$ invariant theories have been
discussed in \cite{Buchmuller:1988wx} (see also \cite{Pitts:2001jw}). The basic idea is to split
the metric degrees of freedom into the determinant $g$, and a new
metric $\hat g_{\mu\nu}= |g|^{-1/n} g_{\mu\nu}$, whose determinant
is fixed $|\hat g|=1$. Note that $\hat g_{\mu\nu}$ is a tensor
density, and under arbitrary diffeomorphisms [for which $
\delta_{\xi} g_{\m\n}=2 \nabla_{(\m}\xi_{\n)}$] it transforms as
\be \delta_{\xi}\hat g_{\mu\nu} = 2 \hat
g_{\lambda(\mu}\hat\nabla_{\nu)} \xi^{\lambda} -{2\over n} \hat
g_{\mu\nu} \hat\nabla_\lambda \xi^\lambda, \ee where $\hat\nabla$
denotes covariant derivative with respect to $\hat g_{\mu\nu}$.
Next, one defines transverse diffeomorphisms as those which
satisfy \be \hat\nabla_\mu \xi^\mu = \partial_\mu \xi^\mu =0, \ee
where in the first equality we have used $|\hat g|=1$. Under such
$\TD$, the new metric transforms as a tensor $$\delta_{\xi}\hat
g_{\mu\nu} = 2 \hat g_{\l(\m}\hat \nabla_{\nu)} \xi^{\l},$$ while
$g$ transforms as a scalar $$\delta_\xi g = \xi^\lambda
\partial_\lambda g.$$ Moreover \cite{Buchmuller:1988wx}, the only
tensors under $\TD$ which can be constructed from $\hat
g_{\mu\nu}$ are the geometric ones, such as
$R_{\mu\nu\rho\sigma}[\hat g]$ and its contractions. It follows
that the most general action invariant under $\TD$ which contains
at most two derivatives of the metric takes the form \be S=\int
\left(-{\chi^2[g,\psi]\over 2} R[\hat g_{\mu\nu}] + L[g,\psi,\hat
g_{\mu\nu}]\right) d^nx.\label{hhh} \ee Here, $\chi$ is a scalar
made out of the matter fields $\psi$ and $g$. Thus, the $\TD$
invariant theories can be seen as ``unimodular" scalar-tensor
theories, where $g$ plays the role of an additional scalar. These
are very similar to the standard scalar-tensor theories, except
for the presence of an arbitrary integration constant in the
effective potential. Following \cite{Buchmuller:1988wx}, we may go
to the Einstein frame by defining $\bar g_{\mu\nu} = \chi^2 \hat
g_{\mu\nu}$, and we have \be S=-{1\over 2}\int \sqrt{-\bar g}\
R[\bar g_{\mu\nu}]\ d^nx + S_M + \int \Lambda \ d^n x,
\label{here}\ee where \be S_M = \int\sqrt{-\bar
g}\left[{(n-1)(n-2)\over 2 \chi^2}\ \bar
g^{\mu\nu}\partial_\mu\chi\partial_\nu \chi +\chi^{-n}
L[\chi,\psi,\bar g_{\mu\nu}]-
\chi^{-n}\Lambda\right]d^nx.\label{there}\ee Here, we have first
eliminated $g$ in favor of $\chi$, and we have then implemented
the constraint $\bar g=\chi^{2n}[g,\psi]$ through the Lagrange
multiplier $\Lambda(x)$. Note that the invariance under full
diffeomorphisms which treat $\bar g_{\m\n}$ as a metric and $\chi$
and $\Lambda$ as scalars is only broken by the last term in
(\ref{here}). In particular, $S_M$ is $\D$ invariant, and since
$\delta_\xi \Lambda = \xi^{\mu}\partial_\mu\Lambda$, it is
straightforward to show that if the equations of motion for
$\psi$, $\chi$ and $\Lambda$ are satisfied, then
$$
|\bar g|^{1/2} \bar \nabla^\mu T_{\mu\nu} = \partial_\mu \Lambda.
$$
Here, we have introduced $T^{\mu\nu} = -2 |\bar g|^{-1/2} \delta
S_M/\delta \bar g_{\mu\nu}$. On the other hand, the Einstein's
equations which follow from (\ref{here}) imply the conservation of
the source $\bar\nabla^\mu T_{\mu\nu}=0$, and therefore we are led
to
$$
\Lambda=const.
$$
This is the arbitrary integration constant, which will feed into
the equations of motion as an extra term in the potential for
$\chi$, corresponding to the last term in Eq. (\ref{there}). In
general, this will shift the height and position of the minima of
the potential for the scalar fields on which $\chi$ depends. In
the particular case where we have $\chi[g,\psi]=1$ in Eq.
(\ref{hhh}), the effect is just an arbitrary shift in the
cosmological constant.

$\D$ invariance is recovered when all terms in $S_M$, given in Eq.
(\ref{there}), except for the last one, are independent of $\chi$.
In that case, $\chi$ is a Lagrange multiplier which sets
$\Lambda=0$, so the freedom to choose the height (or position) of
the minimum of the potential is lost.

Likewise, if the action (\ref{hhh}) does not depend on $g$, then
the symmetry is $\WTD$. The situation is exactly the same as in
the $\TD$ case, where now $\chi=\chi[\psi]$. For instance the
simple action \be S_{\WTD} =-{1\over 2} \int d^n x\ R[\hat
g_{\mu\nu}],\label{ul} \ee which has $\chi=1$, leads to the
equations of motion \be \hat R_{\mu\nu}-\frac{1}{2} \hat R \hat
g_{\mu\nu}= \Lambda \hat g_{\mu\nu},\label{eineq} \ee with
arbitrary integration constant $\Lambda$ (note that in this case
$\hat g_{\mu\nu}=\bar g_{\mu\nu}$). This coincides with the
standard Einstein's equations in the gauge $|g|=1$. The same
action can be expressed in terms of the ``original" metric
$g_{\mu\nu}$ as \be { S}_{\WTD} =-{1\over 2}\int d^n x (-g)^{1/n}
\left(R[g_{\mu\nu}] +{(n-1)(n-2)\over 4n^2}\
\partial^\mu\ln g\
\partial_\mu\ln g\right).\label{nlw} \ee
This is invariant under Weyl transformations \be g_{\mu\nu}\mapsto
\Omega^2(x) g_{\mu\nu},\label{nlwt} \ee since $\hat g_{\mu\nu}$ is
unaffected by these. Of course, it is also invariant under
transverse diffeomorphisms and provides, therefore, an example of
a consistent non-linear completion of a pure spin two Lagrangian
(namely the $\WTD$ Lagrangian which we considered in Section 2),
which is different from GR.

Note that the equations of motion can be derived in two different
ways: directly from (\ref{ul}) under {\em restricted} variations
of $\hat g_{\mu\nu}$ (since by definition $|\hat g|=1$), or from
(\ref{nlw}) under {\em unrestricted} variations of $g_{\mu\nu}$.
Whichever representation is used may be a matter of convenience,
but there seems to be no fundamental difference between the two.
In the latter case, the equations of motion will be completely
equivalent to (\ref{eineq}), although they will only take the same
form in the gauge $|g|=1$.

It is worth mentioning that equations of the form (\ref{eineq})
with an arbitrary $\Lambda$ can also be derived under {\em
unrestricted} variations of an action which is {\em not} invariant
under (\ref{nlwt}). An example is given by \be S=-{1\over 2} \int
\left[\sqrt{-g} R + f(g)\right]d^n x,\label{nonsym} \ee Here, the
second term breaks $\D$ to $\TD$, and there is no Weyl invariance.
However, the equations of motion will give
$$
R_{\mu\nu}-{1\over 2}R g_{\mu\nu}=  \sqrt{-g}\ f'(g)\  g_{\mu\nu},
$$
and from the Bianchi identities it follows that $g$ is an
arbitrary constant (except in the $\D$ invariant case when $f
\propto \sqrt{-g}$ ), a situation identical to (\ref{eineq}). It
is unclear whether the action (\ref{nonsym}) is of any fundamental
significance, since the remaining $\TD$ symmetry does not forbid
an arbitrary function of $g$ in front of $R$, and additional
kinetic terms for $g$. Nevertheless, Lagrangians similar to
(\ref{nonsym}) do arise in the context of certain bigravity
theories where the interaction term between two gravitons breaks
$\D \times \D$ to the diagonal $\D$ times a $\TD$ symmetry
\cite{BDGII}.

To conclude, it should be stressed that it seems to be very
difficult to determine from experiment whether $\D$, $\WTD$ or
just $\TD$ is the relevant symmetry. The difference between $\WTD$
and the rest of $\TD$ theories is just the absence of the extra
scalar. However, this scalar may well have a mass comparable to
the cut-off scale, and in this case it would not be seen at low
energies. Also, at the classical level, the $\WTD$ differs only
from $\D$ in that the cosmological constant is arbitrary. Of
course the measurement of this constant does not reveal too much
about its origin. Therefore, the only ``observable" differences
between both theories may be in the quantum theory
\cite{Alvarezz,Unruh:1988in,Kreuzer,Dragon:1988qf,AlvaVillar,ABGVII}).

\section{Massive fields.}

Let us now turn our attention to the massive case. The most general mass term
takes the form\footnote{Here, we are disregarding the possibility of Lorentz breaking
mass terms, which has been recently considered in \cite{BDGII,Rubakov:2004eb}.}
$$
{\mathcal L}_m = -{1\over 4} m_1^2 h_{\mu\nu}h^{\mu\nu} +{1\over 4} m_2^2 h^2.
$$
First of all, let us note that for $m_1=0$, this mass term is
still invariant under $\TD$. The term $m_2^2 h^2$ gives a mass to
the scalar $h$, but not to the tensor modes $t_{ij}$, which are
traceless. Hence, the analysis of the previous Section remains
basically unchanged. If $-m_2^2
> 0$ is larger than the energy scales we are interested in, the
extra scalar effectively decouples, and we are back to the
situation where only the standard helicity polarizations of the
graviton are allowed to propagate. \footnote{Note also that the
addition of the term $m_2^2 h^2$ to both the $\D$ or the $\WTD$
Lagrangian does not change the propagating degrees of freedom of
the theory. The analogous statement in a non-linear context is
illustrated by the Lagrangian (\ref{nonsym}), where a ``potential"
$f(g)$ is added to a $\D$ invariant Lagrangian (something {\em
does} change, though, by the addition of the potential, since the
new theory does have the arbitrary integration constant
$\Lambda$). Hence, one may in principle construct classical
Lagrangians which propagate only massless spin 2 particles, and
whose symmetry is only $\TD$, although in this case radiative
stability is not guaranteed ({\em i.e.} we may expect other terms,
such as kinetic terms for the determinant $g$, which are not
protected by the symmetry, to be generated by quantum
corrections).}

When $m_1\neq 0$, we must repeat the analysis\footnote{For a
similar analysis in terms of spin projectors see \cite{van}.}.
With the decomposition (\ref{cosdec}), the Lagrangian for the
tensor modes becomes \be
{}^{(t)}{\mathcal{L}}=-\frac{1}{4}t^{ij}\left( \Box+ m_1^2\right)
t_{ij},\label{mtensors}\ee and in order to avoid tachyonic
instabilities we need $m_1^2>0$. For the vector modes, and for
$\beta\neq 1$, the potential term
$$
\Delta {\cal H}_v = {m_1^2\over 2}[\k^2 (F^i)^2 - (V^i)^2],
$$ is added to (\ref{ham}). The contribution proportional to $V^2$
is negative definite. Hence, to avoid ghosts or tachyons we must
take $\beta=1$. In this case, $\dot V^i$ does not appear in the
Lagrangian and $V^i$ can be eliminated in favor of $\dot F^i$.
This leads to \be ^{(v)}{\cal L} = -{1\over 2}\left({\k^2
m_1^2\over \k^2 +m_1^2}\right)\ F^i \left(\Box + m_1^2\right)F^i.
\label{av}\ee Out of the $(N^2+N-2)/2$ polarizations of the
massive graviton in $N+1$ dimensions, $(N^2-N-2)/2$ of these are
expressed as transverse and traceless tensors $t_{ij}$, and $N-1$
are expressed as transverse vectors $F^i$. The remaining one must
be contained in the scalar sector. The scalar Lagrangian can be
written as \be ^{(s)}{\cal L}= ^{(s)}{\cal L}_{\TD} +^{(s)}{\cal
L}_m, \ee where the first term is given by (\ref{scalax}) and the
second is given by \be ^{(s)}{\cal L}_m = -{m_1^2\over 4} (A^2 -2
\k^2 B^2 + N \psi^2 - 2 \k^2 \psi E + \k^4 E^2) +{m_2^2\over
4}(A-N \psi + \k^2 E)^2. \ee Variation with respect to $B$ leads
to the constraint
$$
 m_1^2\ B= {(1-a)(\dot A +\k^2\dot E)-(1-aN) \dot\psi}.
$$
To proceed, it is convenient to eliminate $E$ in favor of the
trace $h$,
$$
\k^2 E = h+ N\psi -A,
$$
and to further express $A$ and $\psi$ in terms of new variables
$U$ and $V$, \bea (N-1)\ A &=& (aN-1)\ h+[2 (N-1) \k^2- N m_1^2]\
U, \nonumber \\ (N-1)\ \psi &=& (a-1)\ h - m_1^2\
(U-V).\label{nv}\eea With these substitutions, and after some
algebra, we find \be ^{(s)}{\cal L}= -{\Delta b\over 4} \dot h^2 +
{[N m_1^2-2(N-1)\k^2]m_1^2 \over 4 (N-1)}\left(\dot V^2- \dot
U^2\right)+{W(h,U,V)\over 4(N-1)^2}, \label{umv}\ee where $\Delta
b$ is given by (\ref{deltab0}) and \bea W \equiv&&
\left\{(N-1)^2(\k^2\Delta b+m_2^2)
-[1+(1-4a+a^2)N+a^2N^2] m_1^2\right\} h^2\nonumber\\
&& +{(N-1)m_1^4\ [(N-2)\k^2-Nm_1^2]}\ V^2\nonumber \\
&& -{m_1^2\ [4(N-1)^2 \k^4 +  (2+N-3N^2) m_1^2 \k^2 + N(N+1)
m_1^4]}\ U^2 \nonumber\\
&&+4(N-1){m_1^2\k^2[N m_1^2 -(N-1) \k^2]}\ UV \nonumber \\
&& +{2 m_1^2\ [(N+1)a-2]}\ [(Nm_1^2-(N-1)\k^2)\ U- (N-1) \k^2\ V]\
h.\label{horta}\eea For $2(N-1)\k^2<Nm_1^2$ the variable $U$ has
negative kinetic energy, whereas for $2(N-1)\k^2> Nm_1^2$ the same
is true of $V$. Thus, the Hamiltonian is unbounded below, unless
\be\Delta b=0.\label{uud}\ee In this case, $h$ is non-dynamical,
and it will implement a constraint between $U$ and $V$ provided
that the coefficient of $h^2$ in $W$ vanishes identically. This
requires \be m_2^2 =\left({1+(1-4a+a^2)N+a^2N^2 \over
(N-1)^2}\right)m_1^2. \label{mrel}\ee As discussed in Section 2,
as long as $a \neq 2/(N+1)$, all kinetic Lagrangians with $\Delta
b=0$ are related to the Fierz-Pauli kinetic term by the field
redefinition (\ref{cov}). Thus, there are only two possibilities
for eliminating the ghost: either the kinetic term is invariant
under $\D$ or it is invariant under $\WTD$.

\subsection{$\D$ invariant kinetic term}

Without loss of generality, we can take $a=b=1$, and from
(\ref{mrel}) we have the usual Fierz-Pauli relation
$$
m_1^2=m_2^2.
$$
Variation with respect to $h$ leads to the constraint \be
(N-1)\k^2 V=[Nm_1^2-(N-1)\k^2] U. \label{babel}\ee In combination
with (\ref{nv}), this yields \be (N-1)\k^2 \psi= m_1^2[N
m_1^2-2(N-1)\k^2]\ U. \label{simp}\ee Substituting (\ref{babel})
in the Lagrangian, and using (\ref{simp}) we obtain
 \be
^{(s)}{\cal L}= -{N \over 4 (N-1)}\ \psi(\Box + m_1^2)\
\psi,\label{as} \ee which is the remaining scalar degree of
freedom of the graviton.

The tensor, vector and scalar Lagrangians
(\ref{mtensors}),(\ref{av}) and (\ref{as}) are not in a manifestly
Lorentz invariant form, and the actual form of the propagating
polarizations is obscured by the fact that the components of the
metric must be found from $F^i$ and $\psi$ with the help of the
constraint equations. Nevertheless, once we know that the system
has no ghosts and all polarizations have the same dispersion
relation, it is trivial to repeat the analysis in the rest frame
of the graviton, $\k=0$. In this frame, the metric is homogeneous
$\partial_i h_{\mu\nu}=0$ and we may write
$$h_{00}={A}, \quad h_{0i}={V}_i, \quad h_{ij}= \psi \delta_{ij} +
t_{ij},$$ where $t^i_i=0$. The Lagrangian for tensors becomes\be
{}^{(t)}{\mathcal{L}}=-\frac{1}{4}t^{ij}\left( \Box+ m_1^2\right)
t_{ij},\label{mtensorss}\ee Vectors contribute to
${\mathcal{L}}^I$ and ${\mathcal{L}}^{II}$, giving \be {}^{({
v})}{\mathcal{L}}= {1\over 2}(\beta - 1) \dot { V}_i^2 + {1\over
2} m_1^2 {V}_i^2, \ee which is non-dynamical in the present case
because $\beta=1$. Likewise, it can easily be shown that the
scalars ${A}$ an $\psi$ are non-dynamical. Therefore, in the
graviton rest frame the propagating polarizations are represented
by the $[N(N+1)/2]-1$ independent components of the symmetric
traceless tensor $t_{ij}$.

\subsection{$\WTD$ invariant kinetic term}

For $a=2/n=2/(N+1)$, the last term in Eq. (\ref{horta})
disappears, and $U$ and $V$ do not mix with $h$. Because of that,
there are no further constraints amongst these variables and the
ghost in the kinetic term in (\ref{umv}) is always present for
$m_1^2\neq 0$. This means that the $\WTD$ theory cannot be
deformed with the addition of a mass term for the graviton without
provoking the appearance of a ghost.

Note that this is so even in the case of a mass term compatible
with the Weyl symmetry, i.e. $m_1^2=n m_2^2$. This relation causes
$h$ to disappear from the Lagrangian, but of course it does
nothing to eliminate the ghost.

\section{Propagators and coupling to matter}

In this section we shall consider the propagators and the coupling
to external matter sources, for the different ``trouble-free"
Lagrangians which we have identified in the previous Sections.

On one hand, we have the standard massless and massive Fierz-Pauli theories,
which have been thoroughly studied in the literature.
There are also the generic ghost-free $\TD$ theories, which satisfy the condition
\be
 \Delta b\equiv b-\frac{1-2a+(n-1)a^2}{n-2} < 0. \label{deltab}
\ee
These may include a mass term of the form $m^2h^2$, which affects the scalar mode
 but does not give a mass to the tensor modes.
The $\WTD$ invariant theory completes the list of possibilities.

Throughout this Section, we will make use of the spin two projector
formalism of \cite{Rivers64}, which is very useful in order to invert the equations of motion.
The properties of these projectors are summarized in Appendix B.

\subsection{Gauge Fixing.}

As noted in \cite{Alvarezz}, for the $\TD$ gauge symmetry there is no linear
covariant gauge fixing condition which is at most quadratic in the momenta.
This is in contrast with the Fierz-Pauli case, where the harmonic condition
contains first derivatives only. The basic problem is that a covariant gauge-fixing
carries a free index, which leads to $n$ independent conditions. This is more than
what transverse diffeomorphisms can handle, since these
have only $(n-1)$ independent arbitrary functions. To be specific,
let us consider the most general possibility linear in $k$,
\be
M_{\a\b\g} h^{\b\g}=0,
\ee
where
\be
M_{\a\b\g}=a_1 \eta_{\a(\b}k_{\g)}+a_2 \eta_{\b\g}k_{\a}.
\ee
In order to bring a generic metric $h_{\mu\nu}$ to this gauge by means of a $\TD$, we have
\be
M_{\a\b\g} h^{\b\g}=M_{\a\b\g} \pd^\b\xi^\g.
\ee
However, deriving the r.h.s. of the previous expression with respect to $x^\a$ and summing in $\a$, this
terms cancels, which implies that
the integrability condition
\be
\pd^\a M_{\a\b\g} h^{\b\g}=0,
\ee
must be satisfied. This simply means that the gauge condition cannot be enforced on generic metrics.

It is plain, however, that the transverse part of the harmonic gauge (which contains only $n-1$
 independent conditions) can be reached by a transverse gauge transformation. The corresponding gauge
fixing piece is  obtained by projecting the harmonic condition with $k^2 \eta_{\mu\nu} - k_\mu k_\nu\equiv
k^2\theta_{\m\n}$:
\be
{\mathcal L}_{gf}=\frac{1}{2 M^4}(\pd_\a\pd^\m\pd^\n h_{\m\n}-\Box\pd^\m h_{\a\m})^2\label{gfterm}
\ee
The gauge fixing parameter is now dimensionful, and this has been explicitly indicated by
denoting it by $M^4$. A study of this kind of term and its associated FP ghosts and
BRST transformations can be found in \cite{AlvaVillar}.
\par
By contrast, in the case of $\WTD$, the additional Weyl symmetry allows
for the use of gauge fixing terms which are linear in the derivatives (such as the standard harmonic gauge).

\subsection{Propagators}

The generic Lagrangian can be written in Fourier space as
\bea
\label{tutto}
&&{\mathcal L}={\mathcal{L}^{I}}+\beta\ {\mathcal{L}^{II}}+ a\
{\mathcal{L}^{III}}+b\ {\mathcal{L}^{IV}}+{\mathcal L}_m+{\cal L}_{gf}={1\over 4}h_{\m\n}K^{\m\n\r\s}h_{\r\s}=\\
&&{1\over 4}h_{\m\n}\Big\{\left(k^2-m_1^2 \right)P_2 +\left[(1-\b)\ k^2-m_1^2+\lambda^2(k)\right]P_1
+ a_s P_0^s+ a_w P_0^w+a_\times P_0^\times\Big\}^{\m\n\r\s}h_{\r\s},\nonumber
\eea
where $P_1$ and $P_2$ are the projectors onto the subspaces of spin 1 and spin 0 respectively, while the
operators $P_0^s$, $P_0^w$ and $P_0^\times\equiv P_0^{sw} + P_0^{ws}$ project onto and mix the different spin 0 components. The definitions and properties of these operators are discussed in Appendix B. The coefficients in front of the spin 0 projectors are given by
\bea
a_s &=& [1-(n-1)b]k^2-m_1^2+ (n-1) m_2^2,\nonumber\\
a_w &=& (1-2\b+2a-b)k^2 - m_1^2 +  m_2^2,\nonumber\\
a_\times&=&\sqrt{n-1}\left[(a-b)k^2+m_2^2\right].
\eea
In (\ref{tutto}), we have included the term $\lambda^2(k) P_1$ which can
be used to gauge fix the $\TD$ symmetry whenever it is present. Indeed, (\ref{gfterm}) can be written as
\be
{\cal L}_{gf}= \lambda^2(k) h_{\mu\nu} P_1^{\mu\nu\rho\sigma}h_{\rho\sigma}.
\label{gf2}
\ee
where $\lambda^2(k)=(1/4M^4)k^6$. Even though we are primarily
 interested in the $\TD$ Lagrangian (which corresponds to $\beta=1$),
we have kept generic $\beta$ throughout this subsection. This can be useful to
handle the cases with enhanced symmetry, since a generic $\beta$ arises, for instance,
from the conventional harmonic gauge fixing term (as we shall see below).
When invertible, the previous Lagrangian yields a propagator $\Delta \equiv K^{-1}$,
\bea
&&\Delta=\frac{P_2}{k^2-m_1^2}+
\frac{P_1}{(1-\b)\ k^2-m_1^2+\lambda^2(k)}+\frac{1}{g(k)}\Big(a_w P_0^s+a_s P_0^w
-a_\times P_0^\times\Big),\nonumber
\eea
where,
\bea
g(k)&=& a_s a_w - a_\times^2.\label{geico}
\eea
Consider a generic coupling of the form
\be
{\cal L}_{int}(x)={1\over 2}(\kappa_1 T^{\mu\nu}+\kappa_2 T \eta^{\mu\nu})
h_{\m\n}\equiv {1\over 2}{\cal T}_{tot}^{\mu\nu}h_{\mu\nu}.
\ee
For conserved external sources,
\be
\pd_\m  T^{\m\n}=0,\label{const}\ee
this coupling is invariant under $\TD$ for all values of $\kappa_1$ and $\kappa_2$.
Moreover, it is $\D$ invariant when $\kappa_2=0$, and $\WTD$ invariant for the special case $\kappa_1=-n\kappa_2$.
The interaction between sources is completely characterized by \cite{Boulware:1973my}
\be
{\cal S}_{int}\equiv {1\over 2}\int d^n k {\cal L}_{int}(k)={1\over 2}\int d^n k \ {\cal T}_{tot}(k)^*_{\mu\nu} \Delta^{\mu\nu\rho\sigma} {\cal T}_{tot}(k)_{\rho\sigma}.
\ee

From the properties of the projectors $P_i$ listed in Appendix B, it is straightforward to show that
\be
{\cal L}_{int}(k)
=\kappa_1^2\  T^*_{\mu\nu}\ \left( \frac{P_2^{\mu\nu\rho\sigma}}{k^2-m_1^2}\right)
\ T_{\rho\sigma}+{\cal P}_0\ |T|^2,\label{flint}
\ee
where the operator
\be
{\cal P}_0={1\over g(k)}\biggl[ {\kappa_1^2 a_w\over (n-1)}
+2\kappa_1\kappa_2 \left(a_w -{a_\times\over\sqrt{n-1}}\right)+\kappa_2^2
\left[ (n-1) a_w + a_s -2 \sqrt{n-1}a_\times\right]\bigg]\label{bigpro}
\ee
encodes the contribution of the spin 0 part. We are now ready to consider
the different particular cases, which we  present by order of increasing symmetry.

\subsection{Massive Fierz-Pauli}

In this case the parameters in the Lagrangian are given by $\beta=a=b=1$
and $m_1^2=m_2^2$. From (\ref{geico}), we have
$$
g(k)=-(n-1)\ m_2^4,
$$
which does not depend on $k$. Because of that, the denominator of the operator ${\cal P}_0$ does not
 contain any derivatives. Its contribution to Eq. (\ref{flint}) corresponds only to contact terms,
 which do not contribute to the interaction between separate sources. We are thus left with the
 spin 2 interaction, which ignoring all contact terms, can be written as
\be
{\cal L}_{int}=\kappa_1^2\ T^*_{\mu\nu}\ \left(\frac{P_2^{\mu\nu\rho\sigma}}{k^2-m_1^2}\right)\ T_{\rho\sigma} = {\kappa_1^2 \over k^2 - m_1^2}\left[T^*_{\mu\nu}T^{\mu\nu}-{1\over (n-1)}|T|^2\right].\label{lintfp}
\ee
The factor $1/(n-1)$ is different from the familiar $1/(n-2)$ which is encountered in linearized GR, and produces
 the well known vDVZ discontinuity in the masless limit \cite{vdvzI,vdvzII,vdvzIII}.

\subsection{$\TD$ invariant theory}

In this case, we set $m_1^2=0$ and $\beta=1$. Note that the gauge fixing term (\ref{gf2}) will not play a role, since the term proportional to $P_1$ does not contribute to the interaction between conserved sources. With these values of the parameters we have
\be
g(k) = (n-2) (\Delta b\ k^2 - m_2^2)\ k^2,
\ee
which is quartic in the momenta. The terms proportional to $\kappa_2$
in the numerator of Eq. (\ref{bigpro}) are also proportional to $k^2$, so this factor drops out and we obtain the propagators
for an ordinary massive scalar particle (provided that $\Delta b < 0$, in agreement with our earlier dynamical analysis).

However, for the first term in Eq. (\ref{bigpro}) (the one proportional to $\kappa_1^2$) there is no global factor of $k^2$ in the numerator, and we must use the decomposition
\be
{1\over g(k)} ={-1\over (n-2)m_2^2}\left({1\over k^2} - {1\over k^2 - {m_2^2\over \Delta b}}\right).
\ee
Substituting in (\ref{bigpro}), and disregarding contact terms, we obtain
\be
{\cal P}_0=-\left({\kappa_1^2 \over (n-1)(n-2)}\right)\ {1\over k^2}-\left(\kappa_2 +{1-a\over n-2} \kappa_1\right)^2{1\over \Delta b k^2 -m_2^2}.\label{wdy}
\ee
Substituting in (\ref{flint}) and adding the contribution of $P_2$ for $m_1^2=0$, which can be read off form (\ref{lintfp}), we have
\be
{\cal L}_{int} = {\kappa_1^2}\left[T^*_{\mu\nu}T^{\mu\nu}-{1\over (n-2)}|T|^2\right]{1 \over k^2}
-\left(\kappa_2 +{1-a\over n-2} \kappa_1\right)^2{|T|^2\over  \Delta b\ k^2 -\ m_2^2}.\label{lintfp2}
\ee
Note that the massless propagator in (\ref{wdy}) combines with the second term in the spin 2 part to give the factor $1/(n-2)$ in front of $|T|^2$. Eq. (\ref{lintfp2}) shows that the massless interaction between conserved sources is the same as in standard linearized General Relativity.

In addition, there is a massive scalar interaction, with effective mass squared
\be
m^2_{eff}= { m_2^2\over \Delta b} >0.
\ee
(note that both parameters $m_2^2$ and $\Delta b$ must be negative, according to our earlier analysis), and effective coupling given by
\be
\kappa^2_{eff} = {-1\over \Delta b}\left(\kappa_2 +{1-a\over n-2} \kappa_1\right)^2.\label{kappaeff}
\ee
These are subject to the standard observational constraints on scalar tensor theories.
If the scalar field is long range, then the strength of
the new interaction has to be very small $\kappa_{eff} \lesssim 10^{-5} \kappa_1$
\cite{Will:2005va,Will:2001mx}.
Alternatively, the interaction could be rather strong, but short range, shielded by a sufficiently large mass $m_{eff}\gtrsim( 30 \mu m)^{-1}$
\cite{Will:2005va,Will:2001mx,Adelberger:2003zx}.

\subsection{Enhanced symmetry}

From general arguments, the interaction between sources in the $\WTD$ theory is expected
to be the same as in standard massless gravity, since both theories only differ by an
integration constant but have the same propagating degrees of freedom.

In fact the result for  $\WTD$ can be obtained from the analysis of the previous
Section by setting $\Delta b=0$. In this case, the term $m_2^2 h^2$ can be thought of as
the additional gauge fixing which removes the redundancy under the additional Weyl symmetry.
With  $\Delta b=0$ the second term in (\ref{lintfp2}) becomes a contact term, and we recover
the same result as in the standard massless Fierz-Pauli theory \cite{Boulware:1973my},
\footnote{Note
 also that the $\WTD$ invariant coupling to conserved sources
 requires $\kappa_1=-n\kappa_2$.  Using this and $a=2/n$ in (\ref{kappaeff})
  we have $\kappa_{eff}=0$, which again eliminates the scalar contribution.}
\be
{\cal L}_{int} = {\kappa_1^2}\left[T^*_{\mu\nu}T^{\mu\nu}-{1\over (n-2)}|T|^2\right]{1 \over k^2},\label{stdfp}
\ee
as expected.

Note that in the $\D$ and $\WTD$ invariant theories,
there is a different possibility for gauge fixing. Rather than using the term (\ref{gf2})
in order to take care of the $\TD$ part of the symmetry, and then the $m_2^2h^2$
 to take care of the Weyl part, we can gauge fix the entire symmetry group with a standard term of the form
\be
{\cal L}_{gf}={\alpha\over 4}\left (\partial_\alpha h^{\alpha\mu}+\gamma \pd^\mu h\right)^2,
\ee
where $\alpha$ and $\gamma$ are arbitrary constants. This can be absorbed in a shift of the parameters $a$, $b$ and $\beta$
$$
a\mapsto a+\alpha \gamma, \quad b\mapsto b-{\alpha\gamma^2\over 2},\quad \beta\mapsto \beta-{\alpha\over 2}.
$$
With these substitutions,
the propagator becomes invertible, even if it is not for the original
values of $a,b$ and $\beta$ which correspond to $\D$ or to $\WTD$. Needless to say,
the result calculated in this gauge coincides with (\ref{stdfp}).

\section{Conclusions}
In this paper we have expanded somewhat the classification of flat
space spin 2 Lagrangians given by van Nieuwenhuizen in \cite{van}.
For the massless theory, we have explicitly shown that unless the
$\TD$ symmetry is imposed, the Hamiltonian is unbounded below and
a classical instability generically develops. We have also
presented in some detail a few potentially interesting theories
which are invariant under this symmetry.
\par
Generic massless $\TD$ theories contain a propagating scalar
proportional to the trace $h$ (note that this is ``gauge
invariant" under $\TD$), which disappears when the symmetry is
enhanced in one of two ways. The standard choice is to consider
the full group of diffeomorphisms $\D$. Another possibility (which
we call $\WTD$) is to impose an additional Weyl symmetry, by which
the action depends only on the traceless part of the metric $\hat
h_{\mu\nu}=h_{\mu\nu}-(1/n)\ h\eta_{\mu\nu}$. This theory is
equivalent to one in which the determinant of the metric $\hat
g_{\mu\nu}=\eta_{\mu\nu}+\hat h_{\mu\nu}$ is kept fixed to unity.
In practice, however, it may be convenient to use the formulation
in which this extra symmetry is present, since it can make the
covariant gauge fixing somewhat simpler. Nonlinear extensions of
the $\TD$ theory have been discussed in \cite{Buchmuller:1988wx},
and they correspond to scalar-tensor theories with an integration
constant. The nonlinear extension of the $\WTD$ theory is a
particular case where the additional scalar is not present.
\par
It is sometimes claimed that General Relativity is the only
consistent theory for spin 2 gravitons. Such discussions
\cite{Ogiev,Deser:1969wk,Boulware:1974sr,Wald:1986bj,Boulanger:2000rq},
however, always assume linearized $\D$ invariance as an input. In
the light of the present discussion, it seems that linearized
$\TD$ or $\WTD$ invariances should be just as good starting
points.

In fact, it may be very difficult to distinguish between the
various options experimentally. The additional scalar may be very
heavy, in which case it can be integrated out leaving no
distinctive traces at low energies. As for the cosmological
constant, it seems clear that we cannot tell from its measurement
whether it corresponds to an integration constant or to a fixed
parameter in the Lagrangian.

An interesting difference between linearized GR and $\TD$
invariant theories is that, as we have shown, the latter cannot be
extended with a mass term for the graviton without provoking the
appearance of a ghost. In this sense, the $\WTD$ theory is more
rigid against small deformations than the standard linearized GR.

It is at present unclear whether $\D$ (rather than $\TD$, or
$\WTD$) is a fundamental symmetry of Nature. Even in string
theory, the connection with GR is on-shell, which does not seem to
exclude the fundamental symmetry from being $\WTD$ (see however
\cite{Ghoshal:1991pu} for a discussion in the context of closed
string field theory). Classically the two theories are almost
identical, but there may be important differences in the quantum
theory
\cite{Unruh:1988in,Kreuzer,Dragon:1988qf,Alvarezz,AlvaVillar}.
These deserve further exploration, and are currently under
research \cite{ABGVII}.

\section*{Acknowledgments}
This work was begun while one of us (E.A.) was a guest at the Departamento di Fisica
 at the University of Roma I (La Sapienza). He is indebted to
 all members of this  group for their wonderful hospitality. J.G.
 thanks Jaume Gomis and Alex Pomarol for useful discussions, and
 the members of the theory group at the University of
Texas at Austin for warm hospitality during the last stages of
this project.
 This work has been partially supported by the
European Commission (HPRN-CT-200-00148) and by research programmes
FPA2003-04597 and FPA2004-04582 (MEC, Spain) and DURSI 2005SGR00082. The work
of D.B. has been supported by FPU2002-0501 (MEC, Spain).

\section*{Appendix A. TDiff lagrangians in terms of gauge invariant quantities.}

As the lagrangian of (\ref{tdl}), ${\mathcal{L}}_{\TD}$,
 is invariant under transverse transformations, one should be able to write it in terms
of invariants under this transformations (for the $\D$ case see e.g.
\cite{Mukhanov:1990me,Bardeen:1980kt,Stewart:1990dy}). It is easy to
see that under a general transformation $h_{\mu\nu} \mapsto h_{\mu\nu}+2\partial_{(\mu}\xi_{\nu)}$ the fields
of the cosmological decomposition
transform as
\bea
t_{ij}&\mapsto& t_{ij},\quad
V_i\mapsto V_i+\partial_0 \xi^T_i,\quad
F_i\mapsto F_i+ \xi^T_i,\quad
A\mapsto A+ 2\partial_0 \xi_0, \nonumber\\
B&\mapsto& B+\partial_0 \eta +\xi_0,\quad
E\mapsto E+2\eta,\quad
\psi\mapsto\psi,\nonumber
\eea
where $\xi_i=\xi_i^T+\partial_i\eta$, with
$\partial^i\xi^T_i=0$. Whereas for a Weyl transformation
$h_{\mu\nu} \mapsto h_{\mu\nu}+ \frac{1}{n}\phi \eta_{\mu\nu}$ only
$A$ and $\psi$ change as
\bea
A\mapsto A+\frac{\phi}{n},\quad
\psi\mapsto \psi -\frac{\phi}{n}.\nonumber
\eea
 For general transverse transformations
the only gauge invariant combinations are
\be
 t_{ij}, \quad w_i=V_i-\partial_0 F_i,
\ee
in the tensor and vector sectors respectively and
\be
\Phi=A-2\partial_0 B+ \partial^2_0
E, \ \psi, \ \Theta=(A-\Delta E),
\ee
for the scalars. In terms of these combinations, the tensor,
vector and scalar part of the lagrangian
(\ref{tdl}) can be written as (we write also the mass term ${\mathcal L}^V=-m^2 h^2$)
\bea
{}^{(t)}{\mathcal{L}}_{\TD}&=&-\frac{1}{4}t^{ij} \Box t_{ij}, \quad
{}^{(v)}{\mathcal{L}}_{\TD}=-\frac{1}{2}
w^i\triangle w^i,\nonumber\\
{}^{(s)}{\mathcal{L}}^I+{}^{(s)}{\mathcal{L}}^{II}&=&\frac{1}{4}\left(-\dot\Theta^2-\Theta\Delta (\Theta-2\Phi)
-2\Delta \psi(\Phi-\Theta)+(n-3)\psi\Delta\psi+(n-1)\dot\psi^2\right),\nonumber\\
{}^{(s)}{\mathcal{L}}^{III}&=&\frac{a}{4}\Big((\Theta-(n-1)\psi)(\Delta(\Theta-\psi-\Phi)-\ddot{\Theta})\Big)
,\nonumber\\
{}^{(s)}{\mathcal{L}}^{IV}&=&-\frac{b}{4}\Big((\dot\Theta -(n-1)\dot \psi)^2+(\Theta-(n-1)
\psi)\Delta (\Theta-(n-1)\psi)\Big)
,\nonumber\\
{}^{(s)}{\mathcal{L}}^V&=&-\frac{m^2}{4}(\Theta-(n-1)\psi)^2,\nonumber
\eea
where $\Delta=\sum_i\partial_i \partial_i=
-\partial^i\partial_i$.
From this decomposition we easily see that $\Phi$ is a lagrange multiplier whose variation yields the
constraint
\be
\label{constr}
\triangle\left((1-(n-1)a)\psi-(1-a)\Theta\right)=0.
\ee

In the case of general diffeomorphisms ($a=b=1$), only two scalar combinations are
gauge invariant, namely $\Phi$ and $\psi$. Thus, the lagrangian for the scalar part
can be expressed as
\be
{}^{(s)}{\mathcal{L}}_{\D}=
\frac{(2-n)}{4}\left(-2\Phi \Delta\psi+(n-1)\dot \psi^2 +(n-3)\psi\Delta \psi \right).
\ee

Concerning the Weyl transformations, we can write only two scalar invariants which
are also scalars for transverse transformations,
\bea
\Xi=\Phi+\psi, \quad \Upsilon=\Theta+\psi.
\eea
Thus, for the Weyl choice $a=\frac{2}{n}$, $b=\frac{n+2}{n^2}$, we can write the lagrangian
as
\bea
\label{kineticconf}
{}^{(s)}{\mathcal{L}}_{\WTD}&=&
\frac{1}{4n^2}\left((n-2)(2n\Xi-(n-1)\Upsilon)\triangle \Upsilon
-(2-3n+n^2)\dot \Upsilon^2\right).
\eea
Thus, varying the lagrangian with respect to $\Xi$ we find the constraint
\be
\Delta \Upsilon=0.
\ee
Besides, the mass term can be written as
\be
{}^{(s)}{\mathcal{L}}^V=-\frac{m^2}{4}(\Upsilon-n\psi)^2.
\ee

\section*{Appendix B. Barnes-Rivers Projectors.}

A useful tool for analyzing the lagrangians involving two component tensors
is provided by the Barnes and Rivers projectors \cite{Rivers64} (see also \cite{van}).
We start with the usual transverse and longitudinal projectors
\bea
&&\theta_{\a\b}\equiv \eta_{\a\b}-\frac{k_{\a}k_{\b}}{k^2}\nonumber\\
&&\omega_{\a\b}\equiv \frac{k_{\a}k_{\b}}{k^2}. \eea
and then define projectors on the subspaces of spin two, spin one, and the two different
spin zero components, labelled by $(s)$ and $(w)$. We introduce also the
convenient operators that map between these two subspaces.
\bea
&&P_2\equiv\frac{1}{2}\left(\theta_{\m\rho}\theta_{\n\sigma}+
\theta_{\m\sigma}\theta_{\n\rho}\right)-\frac{1}{(n-1)}\theta_{\m\n}\theta_{\rho\sigma}\nonumber\\
&&P_0^s\equiv\frac{1}{(n-1)}\th_{\m\n}\th_{\rho\sigma}\nonumber\\
&&P_0^w\equiv\omega_{\m\n}\omega_{\rho\sigma}\nonumber\\
&&P_1\equiv\frac{1}{2}\left(\th_{\m\rho}\omega_{\n\sigma}+\th_{\m\sigma}\omega_{\n\rho}
+\th_{\n\rho}\omega_{\m\sigma}+\th_{\n\sigma}\omega_{\m\rho}\right)\nonumber\\
&& P_0^{sw}\equiv\frac{1}{\sqrt{(n-1)}}\th_{\m\n}\omega_{\rho\sigma},\quad
P_0^{ws}\equiv\frac{1}{\sqrt{(n-1)}}\omega_{\m\n}\th_{\rho\sigma}\nonumber\\
\eea
These projectors obey
\bea
\label{propproj}
&&P_i^a P_j^b=\d_{ij}\d^{ab} P_i^b\nonumber\\
&&P_i^{ab} P_j^{cd}=\d_{ij} \d^{bc} \d^{ad}P_j^{a}\nonumber\\
&&P_i^a P_j^{bc}=\d_{ij}\d^{ab}P_j^{ac}\nonumber\\
&&P_i^{ab} P_j^c = \d_{ij}\d^{bc} P_j^{ac}
\eea
And the traces:
\bea
&&\tr\, P_2\equiv \eta^{\m\n}(P_2)_{\m\n\rho\sigma}=0, \
\tr\, P_0^s =\theta_{\rho\sigma}, \
\tr\, P_0^w=\omega_{\rho\sigma}\nonumber\\
&&\tr\, P_1=0, \
\tr\,P_0^{sw}=\sqrt{n-1}\omega_{\rho\sigma},\
 \tr\, P_0^{ws}=\frac{1}{\sqrt{n-1}}\theta_{\rho\sigma}
\eea

Apart from the previous expressions, these projectors satisfy
\be
P_2+P_1+P_0^w+P_0^s=\frac{1}{2}\left(\d_{\m\n}\d_{\r\s}+\d_{\r\s}\d_{\m\n}\right)
\ee
and any symmetric operator can be written as
\be
K=a_2 P_2+a_1P_1+a_wP_0^w+a_sP_0^s+a_{sw}P_0^\times
\ee
where $P_0^\times=P_0^{sw}+P_0^{ws}$. The inverse of the previous operator
 is easily found from (\ref{propproj}) to be
\be
K^{-1}=\frac{1}{a_2} P_2+\frac{1}{a_1}P_1+\frac{a_s}{a_sa_w-a^2_{sw}}P_0^w+
\frac{a_w}{a_sa_w-a^2_{sw}}P_0^s-\frac{a_{sw}}{a_sa_w-a^2_{sw}}\left(P_0^{ws}+P_0^{sw}\right)
\ee
provided that the discriminant $a_sa_w-a^2_{sw}$ never vanishes.

\end{document}